# Transition metal dichalcogenide nanodisks as high-index dielectric Mie nanoresonators

*Ruggero Verre[+], Denis G. Baranov[+], Battulga Munkhbat, Jorge Cuadra, Mikael Käll[*], Timur Shegai[*]*

Department of Physics, Chalmers University of Technology, 412 96 Göteborg, Sweden

*email: mikael.kall@chalmers.se; timurs@chalmers.se

**Abstract**: Monolayer transition metal dichalcogenides (TMDCs) have recently been proposed as a unique excitonic platform for advanced optical and electronic functionalities. However, in spite of intense research efforts, it has been largely overlooked that, in addition to displaying rich exciton physics, TMDCs also possess a very high refractive index. This opens a possibility to utilize these materials for constructing resonant nanoantennas based on subwavelength geometrical modes. Here we show that nanodisks fabricated from exfoliated multilayer $WS_2$ support distinct Mie resonances and so-called anapole states that can be easily tuned in wavelength over the visible and near-infrared spectral range by varying the nanodisks' size and aspect ratio. We argue that the TMDC material anisotropy and the presence of excitons substantially enrich nanophotonics by complementing traditional approaches based on plasmonics and well-known high-index materials such as silicon. As a proof-of-concept, we demonstrate a novel regime of light-matter interaction, anapole-exciton polaritons, which we realize within a single $WS_2$ nanodisk. Our results thus suggest that nanopatterned TMDCs are promising materials for high-index nanophotonics applications with enriched functionalities and superior prospects.

The possibility of realizing materials that are only a few atoms thick has opened new avenues both for fundamental science and applications. In particular, monolayer TMDC materials with a general formula $MX_2$ (where M = Mo, W; X = S, Se, Te) have attracted significant research attention motivated mostly by their direct bandgaps in the visible and near-infrared part of the spectrum [1]. Each TMDC layer adopts a honeycomb hexagonal lattice that is bound to neighboring layers by weak van der Waals forces, which allows for exfoliation and isolation of individual monolayers [2]. TMDCs have a number of technological advantages, including chemical and thermal stability, high epitaxial quality, compatibility with other materials, and relatively abundance. Due to their unusual exciton physics and large transition dipole moments, TMDCs have been used for strong light-matter interactions [1,3,4], lasing [5], light harvesting and photodetection [6-8], harmonic generation [9,10] and catalytic activity [11]. Furthermore, TMDCs represent a promising class of materials for optical spintronics via spin-selective valley excitations and light emission [12,13]. These effects have resulted in a scientific boom around monolayer TMDCs. However, only a few recent studies have been devoted to multilayer TMDCs [14,15], although their optical properties in the bulk form were extensively studied in the past [16]. In particular, it has been so far disregarded that TMDCs also have a high in-plane refractive index, which exceed $n \approx 4$ and occurs in spite of the highly anisotropic nature of the material.

Large values of the dielectric function have profound implications for visible nanophotonics as it allows the realization of new light matter regimes. Nanoparticles made of high-index dielectric (HID) materials act as nanoantennas that can support multiple geometrical optical resonances, or Mie modes, in the visible to the near-infrared wavelength range. These resonances provide many degrees of freedom for engineering optical responses and for achieving desired functionalities. The interest in all-dielectric nanophotonic has been triggered by their low optical losses and by the fact that standard materials (such as Si, Ge, and GaAs) are compatible with standard CMOS technology [17]. Indeed, a rapidly increasing fraction of contemporary nanophotonics research is devoted to exploring, designing and utilizing HID nanoantennas and metasurfaces and a number of prominent results and proof-of-principle applications have been reported [18,19], including enhancement of nonlinear optical processes [17,20,21] and flat lenses for wavefront manipulation and imaging [22-27]. Thus, introducing novel HID materials with unconventional properties could open up unexplored avenues in nanophotonics research [28].

Here, we demonstrate that nanoantennas fabricated of multilayer $WS_2$, one of the most promising TMDCs for photonics applications, can support distinct multipolar Mie resonances in the visible to near-infrared spectrum. We fabricate WS$_2$ nanodisk antennas with varying aspect ratio and show that the Mie modes disperse with size similarly to previously studied HID nanoantennas made of e.g. Si. Moreover, we find that the antennas support a complex geometrical resonance feature, known as an "anapole" in the near infrared region. We thus introduce a new single crystalline and easily exfoliated material to the library of available HID nanophotonics. Moreover, we also demonstrate that new unusual regimes of light-matter interactions can be observed within a single TMDC nanoantenna by hybridizing Mie resonances with its own material exciton. By varying the nanodisks dimensions, the anapole state is forced to overlap and strongly hybridize with the WS$_2$ A-exciton at around ~615 nm (~2 eV). Analysis based on a simplified coupled mode Hamiltonian implies that the nanoantenna enters the so-called strong coupling regime, that is, the optical excitation oscillates back and forth between the exciton and the geometrical Mie modes. These phenomena are not only of fundamental interest in nanophotonics but could also be of potential use for enhancing light-matter interactions in TMDCs for applications in, for example, light harvesting and photo-catalysis.

**Results:**

The in-plane and out-of-plane components of the relative dielectric constants of bulk WS$_2$ are shown in Fig. 1a [29,30]. The real part of the in-plane permittivity exceeds $\varepsilon_1 \approx 16$, corresponding to refractive index $n = \sqrt{\epsilon} > 4$, in the visible range, decreasing slowly towards the near-infrared, where loss is low. This value is larger than for the most well-studied HID materials in nanophotonics, i.e. silicon and GaAs[28]. The prominent peak at around 615 nm is the so-called A-exciton absorption band due to direct gap transitions at the K-valley of the Brillouin zone. However, due to the van-der-Waals nature of the layer interaction, the out-of-plane permittivity of WS$_2$ is much lower, of the order $\varepsilon_1 \approx 7$. We first investigated how this highly anisotropic dielectric response influences the optical response of a spherical nanoresonator, where shape anisotropy does not occur. We simulated scattering cross-section spectra for a nanosphere of 120 nm radius for the three principal incident wavevector-polarization configurations. All spectra exhibit pronounced resonance peaks, but the peak positions, widths and intensities vary greatly depending on how the nanosphere is excited, despite the spherical geometry (see Fig. 1b). This evidently demonstrates one of the novel features that TMDC provides for high-index nanophotonics, optical anisotropy. To reveal the nature of the spectral features, we tracked the resonance positions as a function of sphere radius (Fig. S1). The results show that the resonances disperse linearly with size above ~700 nm, where material dispersion $\varepsilon(\lambda)$ and absorption is relatively weak. This is the expected behavior for geometrical Mie resonances in high-index nanostructures [31,32]. We also performed multipolar decomposition of the induced polarization inside the nanosphere as a function of illumination conditions (Fig. S2, see Methods for details). This analysis indicates that in-plane polarized incidence excites electric dipole (ED) and magnetic dipole (MD) like modes of the sphere, whereas out-of-plane polarized light only excites a MD-like resonance because of the longer effective wavelength perpendicular to the $WS_2$ planes [33].

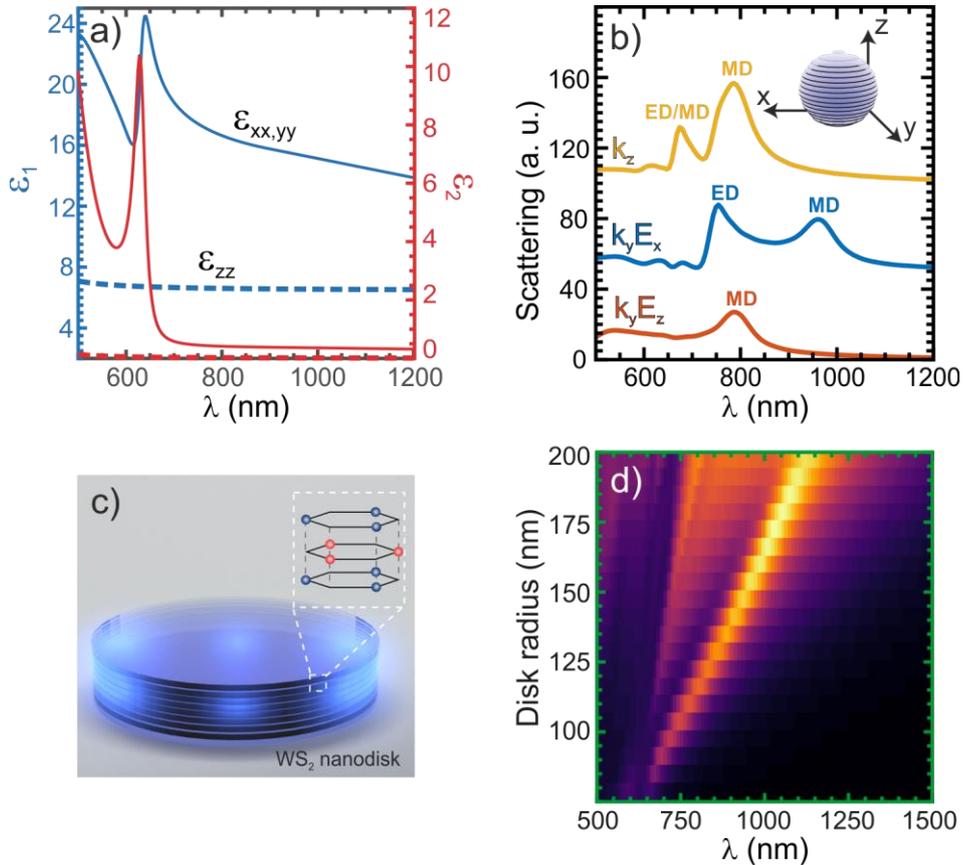

**Figure 1.** (a) In-plane and out-of-plane dielectric function of bulk $WS_2$. (b) Mie resonances of a $WS_2$ nanosphere: The scattering spectra upon the three orthogonal illumination directions for a sphere of 120 nm radius. The electric dipole (ED) and the magnetic dipole (MD) mode character for the various resonances are indicated. (c) Sketch of $WS_2$ nanodisk. (d) Simulated scattering spectra for $WS_2$ nanodisks in air as a function of disk radius for $H$=200 nm.

To experimentally verify Mie resonances in TMDC nanoantennas, we focused on nanodisks because spherical nanoparticles are difficult to realize in practice (see Fig. 1c). We fabricated the nanoantennas from high quality exfoliated multilayer $WS_2$ flakes on Si wafers with a 50 nm thick oxide layer by a combination of electron beam lithography and dry-etching using negative resist as an etching mask (see Methods for details). Figure 1d shows the simulated (FDTD) scattering of $WS_2$ nanodisks as a function of radius for a fixed height of $H$=200 nm and for normally incident light. Similar to the nanosphere case, the nanodisk spectra exhibit distinct geometrical resonances that disperse linearly with size. It is important to note, however, that in this case optical anisotropy plays a smaller role than its shape anisotropy (see Figs. S3-4 for further details on anisotropy effects in nanodisks). Figure 2a shows a typical dark-field (DF) optical microscope image of a sample that contains a wide range of nanodisk sizes. Exfoliated $WS_2$ possess superior crystalline quality and flatness, properties that are highly desirable for nanophotonic applications, and the dry etching process also allows one to obtain nanodisks with high diameter-to-height aspect ratio and vertical side-walls, see scanning electron microscopy (SEM) image in Fig. 2b. The very weak attachment between the substrate and the $WS_2$ makes it is difficult to remove the top resist layer from the nanodisks after processing, but this has negligible influence on the nanodisk optical response (see SI Fig. S6). Micro-Raman characterization of the fabricated nanodisks samples confirms the sample quality through the sharp $WS_2$ phonons at 349 cm$^{-1}$ and 415 cm$^{-1}$ (see Fig. 2c) [34]. Confocal Raman maps demonstrated complete removal of $WS_2$ in the etched region around each nanodisk (Fig. 2c, bottom panels).

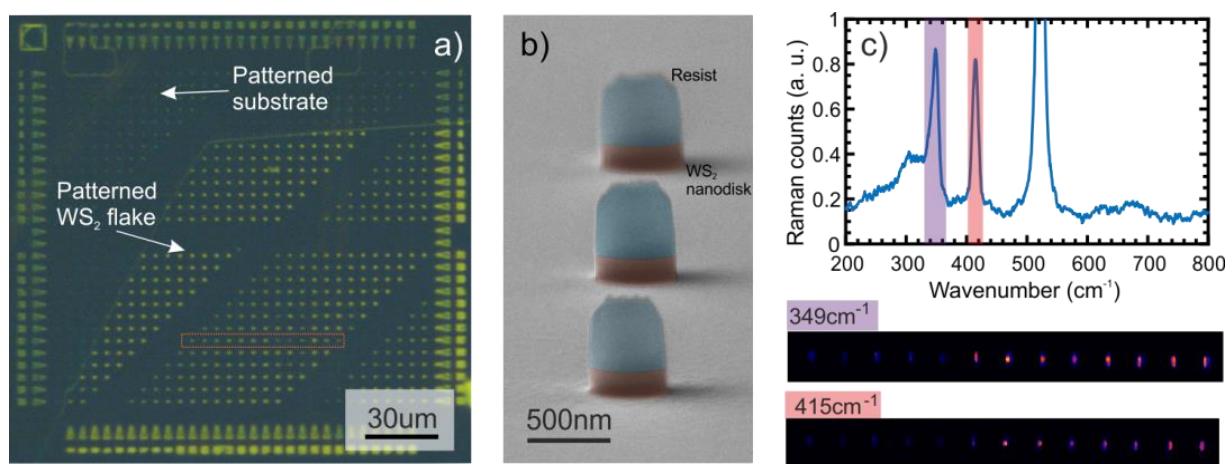

**Figure 2.** (a) Dark field image of the edge of a $WS_2$ flake after nanopatterning. Different contrast between the patterned nanodisks and the substrate is visible. (b) False color side view SEM image of fabricated $WS_2$ nanodisks. (c) Each nanoparticle is composed of single crystalline $WS_2$ as confirmed by the Raman lines at 348 cm$^{-1}$ and 415 cm$^{-1}$ (the intense phonon line at ~530 cm$^{-1}$ comes from the Si substrate). Raman maps at the two wavenumbers for the area indicated in (a) by the red rectangle is also shown.

We measured elastic scattering from individual $WS_2$ nanodisks of varying size using dark-field spectroscopy and then correlated the spectra with morphological data obtained through SEM imaging (Fig. 3a). Figure 3b and 3c show experimental DF scattering spectra and the corresponding FDTD simulations for three nanodisks of 95 nm height and radii of 110 nm, 150 nm, and 200 nm. Experiments and theory are in good overall agreement and both show complex resonance features composed of multiple peaks and dips that disperse with disk size, as anticipated for geometrical Mie modes (see Fig. S5 for additional scattering data for *H*=55, 95, and 150 nm).

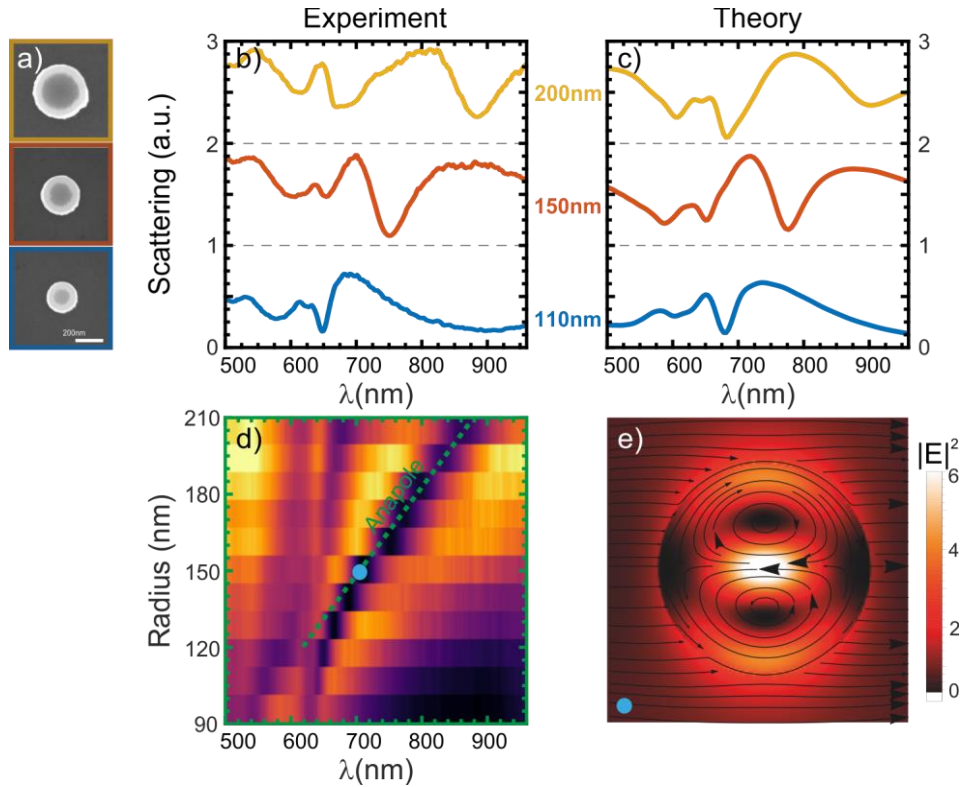

**Figure 3.** (a) Top view SEM images of $WS_2$ nanodisks with the different radii and *H*=95 nm. (b) Experimental and (c) simulated DF scattering spectra of the corresponding $WS_2$ nanodisks. Several peaks and dips are observed and related to geometrical resonances. (d) Color-coded measured DF scattering spectra for $WS_2$ nanodisks of *H*=95 nm and varying radii. The dip at long wavelength, showing a linear dispersion, is attributed to the anapole indicated by the green dashed line. (e) The simulated spatial distribution of the electric field intensity and field lines across the center of the nanodisk in the horizontal plane at the scattering dip for *R*=150 nm revealing the typical anapole-like pattern.

The most spectacular feature in the scattering data is clearly the deep and asymmetric dip that disperses linearly with *R* between ~650 and 900 nm (Fig. 3d). It is now known that certain nanoparticles, in particular HID nanodisks with high diameter-to-height aspect ratio, can support a so-called "anapole" interference characterized by vanishingly small elastic scattering and at the same time very high field-enhancement in the interior of the nanoparticle [35-37]. This behavior originates from destructive (out-of-phase) interference between electric and toroidal dipolar charge oscillations with identical amplitude and far-field patterns but drastically different radiative rates [35,38-40]. In this sense, the anapole is analogous to destructive Fano interferences between overlapping broad dipole-allowed and narrow dipole-forbidden modes observed in many plasmonic systems [41,42]. In order to test if the observed scattering dip can be assigned to an "anapole", we performed multipole decomposition of the internal polarizations in the nanodisks. The resulting expansions in electric and toroidal Cartesian dipole moments indeed points towards an "anapole" origin, although the amplitudes of the two modes do not perfectly coincide at the position of the scattering dip (see Fig. S7). Nevertheless, the calculated field distribution inside the disk (Fig. 3e) exhibits the distinct anapole-like pattern known from previous studies of this phenomenon in e.g. Si nanodisk [36]. We note that an ideal anapole cannot arise in the $WS_2$ system due to dissipative losses, which prevent complete suppression of scattering according to the optical theorem [43].

$WS_2$ nanodisks thus support various multipolar geometrical modes, as well as non-scattering (anapole) states. We will now demonstrate that these resonances can couple to and hybridize with the exciton resonance of the same $WS_2$ nanodisk. Strong coupling is an optical phenomenon where energy transfer between different resonant modes of a system occurs on time scales faster than any dissipation, which leads to the vacuum Rabi splitting of the resonances and the emergence of hybrid polaritonic dressed states [44,45]. These hybrid states have attracted considerable attention recently as they are of interest for quantum and nonlinear optics applications, as well as for modification of photochemical reactions [46-49]. Mie resonances of Si and perovskite nanostructures, as well as $WS_2$ nanotubes, were recently shown to hybridize within various excitonic systems [50-54]. Here we observe hybridization of anapoles and excitons within a single $WS_2$ nanodisk.

We first demonstrate coherent exciton-anapole coupling by means of FDTD simulations for a nanodisk with $H$=55 nm and $R$=128 nm (see Fig. 4a). To illustrate the effect of coupling, we incorporated an artificial dielectric function that accounts either for the background refractive index of $WS_2$ or its A-exciton, respectively (see Methods and Fig. S8-S9). For the case of a hypothetical lossless material with the A-exciton transition artificially switched off, we observe the anapole dip (see green curve in Fig. 4a). If, on the contrary, one uses an artificial permittivity including only the A-exciton, but not the high background permittivity, a single scattering peak at the exciton wavelength is observed (red curve). When the coupling is "switched on" by using a permittivity that includes both the exciton resonance and the high background index, the two spectral features hybridize leading to the emergence of two scattering dips at 580nm and 680nm (purple curve). Besides the two dips, one can see the third small dip in between, which originates from absorption by uncoupled excitons.

To experimentally realize such peculiar regime of interaction, we focused on the $H$=55 nm data set, which exhibits fewer high-order multipole resonances than the larger disks and therefore shows cleaner coupling behavior. The resulting DF scattering map for $H$=55 nm nanodisks (Fig. 4b) reveals a rich collection of resonances well reproduced by FDTD simulations (Fig. 4c-d). Overall, the scattering presents a broad background, on top of which two dips can be clearly identified – one associated with absorption by the $WS_2$ exciton at 630 nm, and the other one that scales linearly with radius for large disks and is associated with the anapole. Remarkably, for the nanodisk radius in the vicinity of 130 nm, the scattering signal exhibits the desired anti-crossing. The experimentally measured anticrossing behavior allows to estimate the vacuum Rabi splitting between the two anapole-exciton modes on the order of $\Omega \approx 190$ meV. This value is larger than the respective line-width of the dips, confirming strong coupling between the two resonances. FDTD calculations also show anti-crossing in absorption (see Fig. 4c), which provides further evidence for the strong coupling regime reached between anapole and exciton in this case [55].

To elaborate further on the strong coupling scenario, we employ a simple Hamiltonian model based on the temporal coupled mode theory [56]. The model describes the interaction between two Mie modes of the nanodisk (bright and dark one), whose destructive interference results in the anapole dip, and the A-exciton of the material (see Section S3.2). In order to apply the model, we fitted the calculated FDTD scattering spectra with the analytical expression for the scattered intensity according to the Hamiltonian model (see section S3.3 for details of this procedure). The so obtained eigen energies and coupling constants of the system produce scattering spectra that closely resemble the experimental and FDTD data (Fig. 4f). This model illustrates the coupling process in the system: first, incident light excites the system by radiative coupling with the bright state. Scattering by these modes at the anapole wavelength is cancelled due to their destructive interference in the far field. The energy stored inside the $WS_2$ nanodisk can now be transferred to the exciton through the near-field (direct) coupling. In the frequency domain, this picture gives rise to two dips in scattering, one associated with the destructive far-field interference, and the other one associated with the A-exciton non-radiative decay and these two features exhibit anti-crossing behavior.

Finally, we use the Hamiltonian model to verify that the anapole-exciton interaction reaches the strong coupling regime by simulating the transient dynamics of the exciton amplitude following an initial excitation. We adopt the parameters extracted by the fitting procedure for the $R$=128 nm nanodisk, which exhibits near-zero anapole-exciton detuning, and solve the dynamical equations governing the mode amplitudes (section S3.1) with the exciton being excited at $t$=0. The resulting amplitudes clearly exhibit Rabi oscillations with a period of around 12 fs due to strong coupling occurring mostly between the exciton and the dark mode (Fig. 4e).

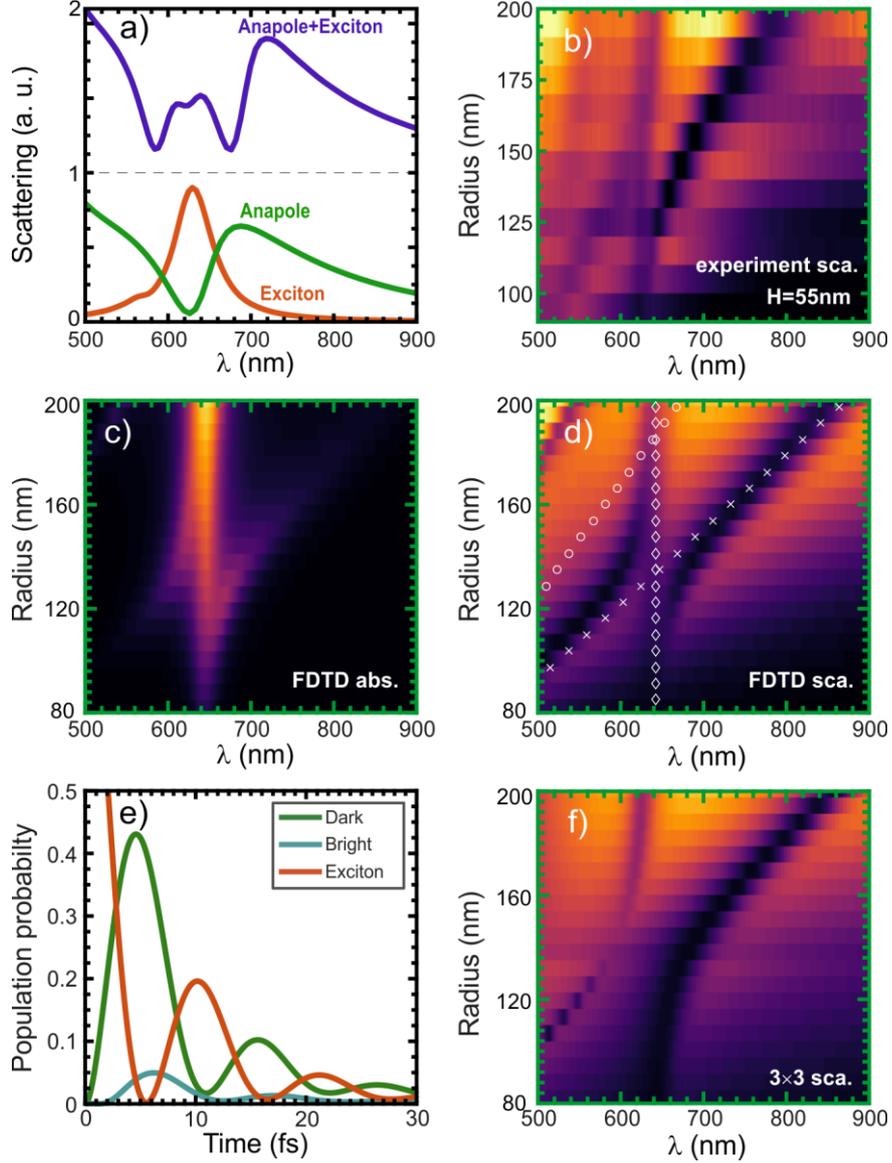

**Figure 4.** (a) Calculated scattering spectra for nanodisk of $H$=55 nm and $R$=140 nm in air with exciton resonance turned off (green), high-index background turned off (red), and actual permittivity of bulk $WS_2$ (blue). (b) Experimental DF spectra for thin ($H$=55 nm) nanodisks with varying diameter show anti-crossing behavior between the two dark states. (c-d) FDTD calculated absorption (c) and scattering (d) maps for nanodisks of H=55 nm and variable radii in air using the artificial permittivity combining the high background index and the exciton resonance. Clear anti-crossing in both absorption and scattering is observed. In (d) the white symbols are uncoupled eigen energies extracted using the 3×3 Hamiltonian representation of the coupling problem. (e) Rabi oscillations between the dark, bright and exciton resonances of the $WS_2$ nanodisk of *H*=55 nm and *R*=128 nm calculated with the use of the coupled mode theory. (f) The scattering intensity calculated with the 3×3 Hamiltonian model using the eigen energies and coupling constants, obtained by fitting the corresponding FDTD spectra.

**Discussion and conclusion:**

In conclusion, we have introduced the concept of TMDC nanoantennas and demonstrated that $WS_2$ nanodisks can be used as a new platform for high-index nanophotonics. Experimental data reveal a rich family of complex resonant modes, including the so-called "anapole" of high current interest. Moreover, TMDC materials possess unique characteristics that allow one to investigate optical regimes only rarely encountered in the context of nanophotonics, for example the effect of an anisotropic dielectric response in a spherically symmetric nanoparticle. Importantly, we also found that the dimensions of a single $WS_2$ nanodisk can be tuned to generate an unusually strong coupling between its geometrical optical modes and the intrinsic excitons of the TMDC material itself. The resulting polaritonic regime is achieved in a compact photonic system composed of a single nanoantenna. More generally, it is worth noting that the TMDC family includes a large class of materials, such as $WSe_2$, $MoS_2$, $MoSe_2$, $NbSe_2$ and their van-der-Waals heterostructures, with similar optical properties but potentially unique physical, chemical and material characteristics. This provides a rich playground in nanophotonics that we anticipate will result in many more exciting observations in the near future.

**Acknowledgements**

We would like to acknowledge the financial support from Knut and Alice Wallenberg Foundation, Chalmers Excellence Initiative Nano and the Swedish Research Council (Vetenskapsrådet).

**Corresponding Author**

* Email: mikael.kall@chalmers.se, timurs@chalmers.se

[+]These authors contributed equally to this work